\documentclass[aps,12pt]{revtex4}
\usepackage[dvips]{graphicx}
\begin{document}

\title{Comment on ''Low-dimensional spin $S$=1/2 system at the quantum
critical limit: Na$_{2}$V$_{3}$O$_7$'' }
\author{R.J. Radwanski}
\homepage{http://www.css-physics.edu.pl} \email{sfradwan@cyf-kr.edu.pl}
\affiliation{Center for Solid State Physics, S$^{nt}$Filip 5, 31-150 Krakow, Poland,\\
 Institute of Physics, Pedagogical University, 30-084 Krakow, Poland}
\author{Z. Ropka}
\affiliation{Center for Solid State Physics, S$^{nt}$Filip 5, 31-150 Krakow, Poland}
\pacs{75.20.H\\}
\maketitle
From the experimentally measured temperature dependence of the magnetic susceptibility $\chi (T)$ of
Na$_{2}$V$_{3}$O$_{7}$, exhibiting at temperatures below 100 K drastic violation of the Curie-Weiss law, Gavilano {\it
et al.} \cite{1} have inferred that the effective moment of the V$^{4+}$ ion in Na$_{2}$V$_{3}$O$_{7}$ is reduced by
the one order of magnitude upon reducing the temperature from 100 to 10 K. In the figure 1 of Ref. \cite{1} it is seen
that after taking into account the diamagnetic contribution $\chi _{o}$ (= -0.0005 emu/mol f.u.)) the inverse
susceptibility shows in the temperature range 100-300 K a straight line behavior with the effective moment p$_{eff}$ of
1.9 $\mu _{B}$ per V ion. Another straight line between 20 and 1.9 K implies $p_{eff}$ of one order of magnitude
smaller. Gavilano {\it et al.} provide an explanation that ''The reduction of the effective magnetic moment is most
likely due to a gradual process of moment compensation via the formation of singlet spin configurations with most but
not all of the ions taking part in this process. This may be the result of antiferromagnetic interactions and
geometrical frustration.'' They further conjectured, recalling the structural speciality (nanotubes), ''the
compensation of eight out of the nine V spins ...'' in order to reproduce the observed reduction of the effective
moment by one order of magnitude. Moreover, they also found that Na$_{2}$V$_{3}$O$_{7}$ shows no sign of the magnetic
order down to 1.9 K - we find this experimental observation to be in sharp contradiction with the earlier conclusion
about the presence of strong antiferromagnetic interactions.

In this Comment \cite{2} we would like to argue that this drastic violation
of the Curie-Weiss law can be understood as caused by well-known
conventional phenomena like the crystal field (CEF) interactions but with
taking into account the intra-atomic spin-orbit (s-o) coupling for the V$%
^{4+}$ ion (3$d^{1}$ configuration). There is the general agreement that in
Na$_{2}$V$_{3}$O$_{7}$ there are V$^{4+}$ ions. Calculations, motivated by
Gavilano {\it et al}.'s paper, of the electronic structure of the V$^{4+}$
ion (3$d^{1}$ configuration) under the action of the CEF and s-o
interactions with the resulting $\chi (T)$ dependences are presented in Ref. %
\cite{3}. These calculations followed our earlier results \cite{4,5,6}. Refs %
\cite{5} and \cite{6} have been devoted to the 3$d^{1}$ configuration. Our
calculations prove the enormous influence of the spin-orbit coupling and of
crystal-field interactions on the temperature dependence of the paramagnetic
susceptibility (and other thermodynamical properties) revealing the drastic
violation of the Curie-Weiss law at low temperatures. We think that our
results are quite obvious for some people, in particular for those having
experience with CEF effects in rare-earth compounds and knowing a
70-year-old book of Van Vleck \cite{7}. This Comment is motivated by the
subsequent rejection by the Editor of Phys. Rev. Lett. of our paper
''Spin-orbit origin of large reduction of the magnetic moment in in Na$_{2}$V%
$_{3}$O$_{7}$'' \cite{3}, in which we propose our CEF+s-o-based explanation.

In Fig. 1 we present the calculated influence of the spin-orbit coupling and
crystal-field interactions on the atomic-scale susceptibility. From this
figure the significant departure from the Curie law as well as from the $S$%
=1/2 behavior due to the spin-orbit coupling and distortions is clearly
seen, in particular at low temperatures. Curve 3, calculated for the
octahedral crystal field $B_{4}$=+200 K, the spin-orbit coupling $\lambda
_{s-o}$= +360 K and an off-octahedral trigonal distortion $B_{2}^{0}$=+9 K,
reproduces very well measured experimental data (shown as x in Fig. 1, after
Refs \cite{1,8}) with taking into account the diamagnetic term $\chi _{o}$
of -0.0007 $\mu _{B}$/T V-ion ($\simeq $ -0.0004 emu/mol V)). We treat this
coincidence as not fully relevant owing to the much more complex local
symmetry of the V$^{4+}$ ion in Na$_{2}$V$_{3}$O$_{7}$, to a large
uncertainty in the evaluation of the diamagnetic term and of the
paramagnetic susceptibility measured on a polycrystalline sample. We take,
however, the reached agreement as strong argument for the high physical
adequacy of our CEF+s-o approach and as strong indication for the existence
of the fine electronic structure in Na$_{2}$V$_{3}$O$_{7}$, originating from
the V$^{4+}$ ion, determined by crystal-field and spin-orbit interactions.
Very important is the fact that our approach is able to reproduce the
overall $\chi (T)$ dependence in the full measured temperature range and
that it reproduces the absolute value (it is really not trivial result) of
the macroscopic magnetic susceptibility. For the recalculation of the
microscopic atomic-scale susceptibility we take into account only the number
of the V ions involved - here for the molar susceptibility simply the
Avogadro number is taken.
\begin{center}
\begin{figure}[ht]
\includegraphics[width = 13 cm]{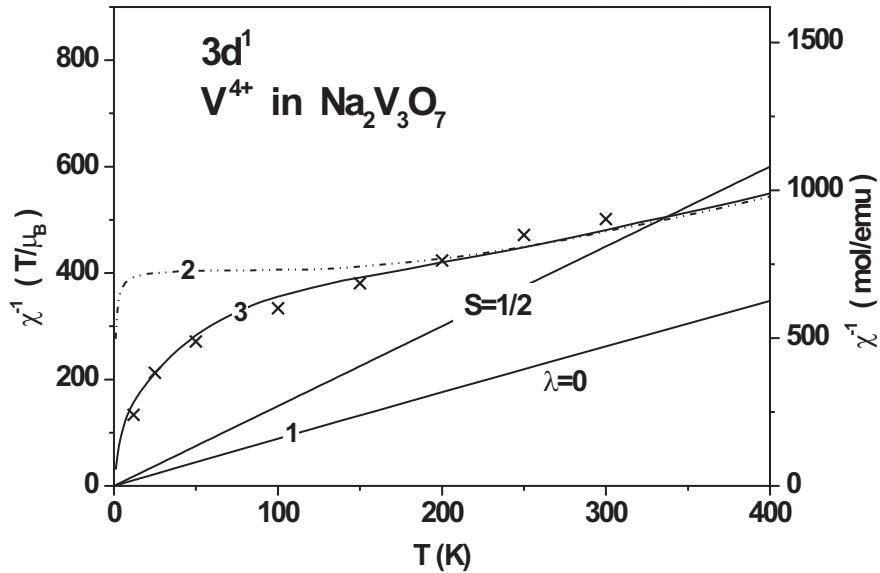}
\caption{The calculated temperature dependence of the atomic-scale paramagnetic susceptibility shown in the $\chi
^{-1}$ {\it vs} $T$ plot for the 3$d^{1}$ configuration in the V$^{4+}$ ion for different physical
situations: \ line (1) - for the purely octahedral crystal field with $B_{4}$%
=+200 K ($\lambda _{s-o}$= 0); line (2) - the octahedral crystal field with $%
B_{4}$=+200 K in combination with the spin-orbit coupling $\lambda _{s-o}$= +360 K; line (3) shows the influence of the
off-octahedral trigonal distortion $B_{2}^{0}$=+9 K. Curve (3) reproduces very well measured experimental data (x,
after Refs \cite{1,8} with taking into account the diamagnetic term $\chi _{o}$ of -0.0007 $\mu _{B}$/T V-ion ($\simeq
$ -0.0004 emu/mol V)).}
\end{figure}
\end{center}

The ground state has weak magnetic moment due to the large orbital moment compensating the spin moment. In
Na$_{2}$V$_{3}$O$_{7}$ there is a low-energy electronic structure originating from the V$^{4+}$-ion fine structure like
that shown in Fig. 1(4) of ref. \cite{3}. The energy level scheme contains 2 close excited doublets at 58 \ and 580 K
and the
Kramers-doublet ground-state moment amounts to $\pm $0.21 $\mu _{B}$ (=2$%
\cdot $($\pm $0.23)$\mp $0.25). It is important to realize that whatever
lower symmetry is in case of the V$^{4+}$ ion 5 Kramers doublets always are.
The used parameters $B_{4}$ (+200 K), $\lambda _{s-o}$ (+360 K $\simeq $ 31
meV) and $B_{2}^{0}$ (+9 K) have clear physical meaning.

These both theoretical approaches can be experimentally distinguished. In
the approach of Ref. 1 only 1/9 vanadium atoms contribute to the
susceptibility at low temperatures whereas in the QUASST\ approach all
vanadium atoms basically equally weakly contribute to $\chi $. But for
searching for the truth the open exchange of information and the open
discussion must be gwaranteed. Moreover, our approach allows to calculate a
number of zero-temperature properties, the low-energy electronic structure
and other thermodynamical properties like has been presented in Refs \cite%
{5,6,7,9,10,11,12,13}. All of them can be experimentally verified, the
low-energy electronic structure by the EPR\ spectroscopy for instance, in
contrary to explanation of Ref. \cite{1}. We hope that this discussion will
help to solve the very serious scientific problem about the role played by
the spin-orbit interactions and by orbital magnetism in 3$d$-atom containing
compounds.

In conclusion, we argue that well-established phenomena like the spin-orbit
coupling, even quite weak, crystal-field interactions of low symmetry and
the orbital magnetism have to be taken into consideration for explanation of
non-trivial electronic and magnetic properties of Na$_{2}$V$_{3}$O$_{7}$
before other exotic mechanisms are proposed.

\end{document}